\documentclass[journal,10pt]{IEEEtran}
\usepackage{graphicx, epstopdf}
\usepackage{amssymb, amsmath}
\usepackage{setspace}
\usepackage{acronym}
\usepackage{mathrsfs}
\usepackage{ifthen}
\usepackage{textcomp}
\usepackage{float}
\usepackage{subfigure}
\usepackage[table]{xcolor}
\usepackage{color,soul}
\usepackage{cite}
\usepackage{setspace}
\usepackage{mathtools}
\usepackage{bm}
\usepackage{url}
\usepackage[switch]{lineno}

\newcommand{\eqapprox}[1]{e^{\{#1\}}}
\newcommand{\veqapprox}[1]{\mathbf{e}^{\{#1\}}}
\newcommand{\waverage}{w_\text{av}}

\newcommand{\win}{w}
\newcommand{\dout}{d^{\text{out}}}

\IEEEoverridecommandlockouts
\begin{document}

\title{Node-Level Resilience Loss in \\ Dynamic Complex Networks}


\author{Giannis Moutsinas\textsuperscript{1}, Weisi Guo\textsuperscript{1,2*}

\thanks{\textsuperscript{1}School of Engineering, University of Warwick, United Kingdom. \textsuperscript{2}The Alan Turing Institute, United Kingdom.} }

\markboth{current preprint version (v1) August 2018}
{Submitted paper}
\maketitle

\begin{abstract}
In an increasingly connected world, the resilience of networked dynamical systems is important in the fields of ecology, economics, critical infrastructures, and organizational behaviour. Whilst we understand small-scale resilience well, our understanding of large-scale networked resilience is limited. Recent research in predicting the effective network-level resilience pattern has advanced our understanding of the coupling relationship between topology and dynamics. However, a method to estimate the resilience of an individual node within an arbitrarily large complex network governed by non-linear dynamics is still lacking. Here, we develop a sequential mean-field approach and show that after 1-3 steps of estimation, the node-level resilience function can be represented with up to 98\% accuracy. This new understanding compresses the higher dimensional relationship into a one-dimensional dynamic for tractable understanding, mapping the relationship between local dynamics and the statistical properties of network topology. By applying this framework to case studies in ecology and biology, we are able to not only understand the general resilience pattern of the network, but also identify the nodes at the greatest risk of failure and predict the impact of perturbations. These findings not only shed new light on the causes of resilience loss from cascade effects in networked systems, but the identification capability could also be used to prioritize protection, quantify risk, and inform the design of new system architectures.
\end{abstract}

\begin{figure}[t]
    \centering
    \includegraphics[width=1.00\linewidth]{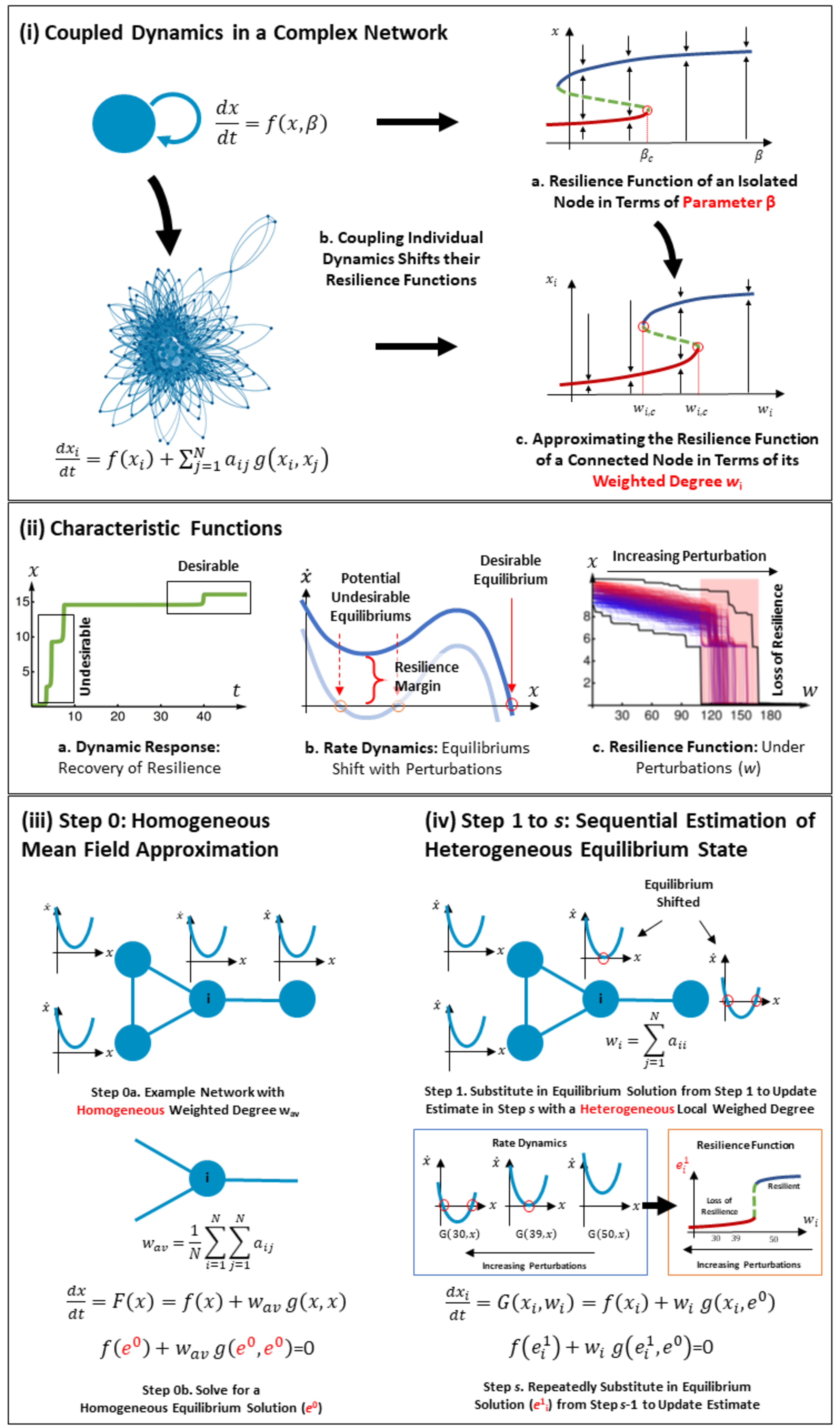}
    \caption{{\bf Estimating Node Level Resilience in Complex Networks.} \textbf{(i)} Problem Definition - coupled dynamics in a complex network, where each node is governed a self-dynamic $f(.)$ and a coupling dynamic $g(.)$. This is connected together through a complex network. Individual resilience is sensitive to $\beta$, and connected resilience is sensitive to the topological measure $w_{i}$. \textbf{(ii)} Characteristic Functions - (a) \textbf{dynamic response} $x(t)$ shows how a system or node can recover to desirable values (resilience behaviour), (b) \textbf{rate dynamics} $\dot{x}(x)$ gives desirable and potentially undesirable equilibrium solutions that changes with perturbations, and (c) \textbf{resilience function} $x(w)$ describes how perturbations that change network property $w$ causes a loss of $e$, leading to unrecoverable collapse (loss in resilience). \textbf{(iii)} Estimation Algorithm - Step 1: mean field approximation using weighted degree to estimate the \textbf{homogeneous} equilibrium solution $e^{0}$ at all nodes. \textbf{(iv)} Estimation Algorithm - Step 2 to $s$: sequential substitution of equilibrium solution $e^{s-1}_{i}$ into $x_{j}$ to estimate next \textbf{heterogeneous} equilibrium solution $e^{s}_{i}$.}
    \label{fig:1}
\end{figure}

\section*{Significance Statement}
A gap in understanding exists between individual dynamics and the coupled dynamics of a large-scale networked complex system. Here, we present a framework for tractably analyzing the resilience of individual nodes as a function of the individual dynamics and the network property. Quantifying connected resilience as a function of the dynamics enables us to prioritize actions more effectively and predict resilience loss more accurately. Conversely, we may also discover hidden cascade effects, whereby disconnecting a weakly connected node can lead to failure in other nodes. In general, the node-level precision methods developed here will enable practitioners in ecology, infrastructure, and other application areas to prioritize protection and intervention resources, such as maintenance, preservation, rewiring, and upgrades.

\section{Introduction}

Organized behaviour in economics\cite{Bardoscia17}, infrastructure\cite{Zimmerman11}, ecology\cite{Lu16}, biology\cite{Wilhelm04}, and human society\cite{Ellinas17} often involve large-scale networked systems, coupling together relatively simple dynamics to achieve complex behaviour. A critical part of the organized behaviour is the ability for a system to be resilient - the ability to retain original functionality after a perturbation \cite{May72} or failure. When failures lead to disconnections, traditional robustness measures only consider topological changes, e.g. random removals to giant component collapse \cite{Callaway00}. Yet, we know that the dynamics can play an important role, and often systems fail long before they are disconnected, e.g. connected components can lose desirable functionality due to cascade effects. Whilst the precise form of resilience loss is different for each ecosystem, what they have in common is the incapacity to bounce back to the original desirable state. 

\subsection{Background}

For the example illustrated in Fig.\ref{fig:1}i, a change in circumstance (represented by $\beta$ in Eq.\eqref{eq:f}) can shift behaviour from a desirable (blue) to an undesirable (red) state. The system cannot bounce back to this desirable state and this is defined as a loss in resilience. Over the last few decades, practitioners have built up a strong understanding of each individual subsystem’s functional resilience. For example, a simple one-dimensional subsystem can be described by how parameter $x$ changes:
\begin{equation}
\frac{dx}{dt} = f(x,\beta), 
\label{eq:f}
\end{equation} where at equilibrium $f(x=e,\beta)=0$ and $\frac{d f}{dx } |_{x=e} < 0$ maps to the \textbf{resilience function} $x(\beta)$ given in Fig.\ref{fig:1}i-a. 

Whilst many physical, biological, ecological, social, and engineering systems have subsystems that can be described by Eq.\eqref{eq:f}, we do not have tractable understanding of the resilience function in large-scale networked dynamics (see Fig.\ref{fig:1}i-b):
\begin{equation}
\frac{dx_{i}}{dt} = f(x_{i}) + \sum_{j}^{N}a_{ij}g(x_{i},x_{j}), 
\label{eq:fg}
\end{equation} where each subsystem (node) $i$'s behaviour is described by a self-dynamic $f(\cdot)$ and a coupling dynamic $g(\cdot)$ with node $j$ via the connectivity matrix $A_{ij}$. In general, we do not know very well how functional resilience maps to the topological resilience (e.g. properties of $A_{ij}$) of connected ecosystems. Indeed, recent research have begun to address this by mapping the overall effective dynamics of a networked system to its topological structure and individual dynamics \cite{Barzel13, Gao16}: $\dot{x}_{\text{eff}}(\beta_{\text{eff}},x_{\text{eff}})$, where $x_{\text{eff}}$ yields the effective mean network dynamics and $\beta_{\text{eff}}$ captures effective aspects of the network topology. This work has been extended to consider negative interactions \cite{Tu_PRE_17}, noise effects \cite{Liang17}, attack strategies \cite{Lv17}, and applied to critical infrastructure areas \cite{Nitzbon17}. Other network level predictions using dimension reduction techniques have also been developed to yield similar insights \cite{Jiang_PNAS_17}. However, we still do not understand the resilience and dynamics of individual nodes. As we will show, many systems can exhibit a common network-level effective dynamic, but have different node level dynamics. The precision to identify the resilience function at the node-level is sorely needed in all application domains in order to inform ground operations (e.g. prioritize conservation in ecology, enhance monitoring in infrastructures).    
\begin{figure}[t]
    \centering
    \includegraphics[width=1.00\linewidth]{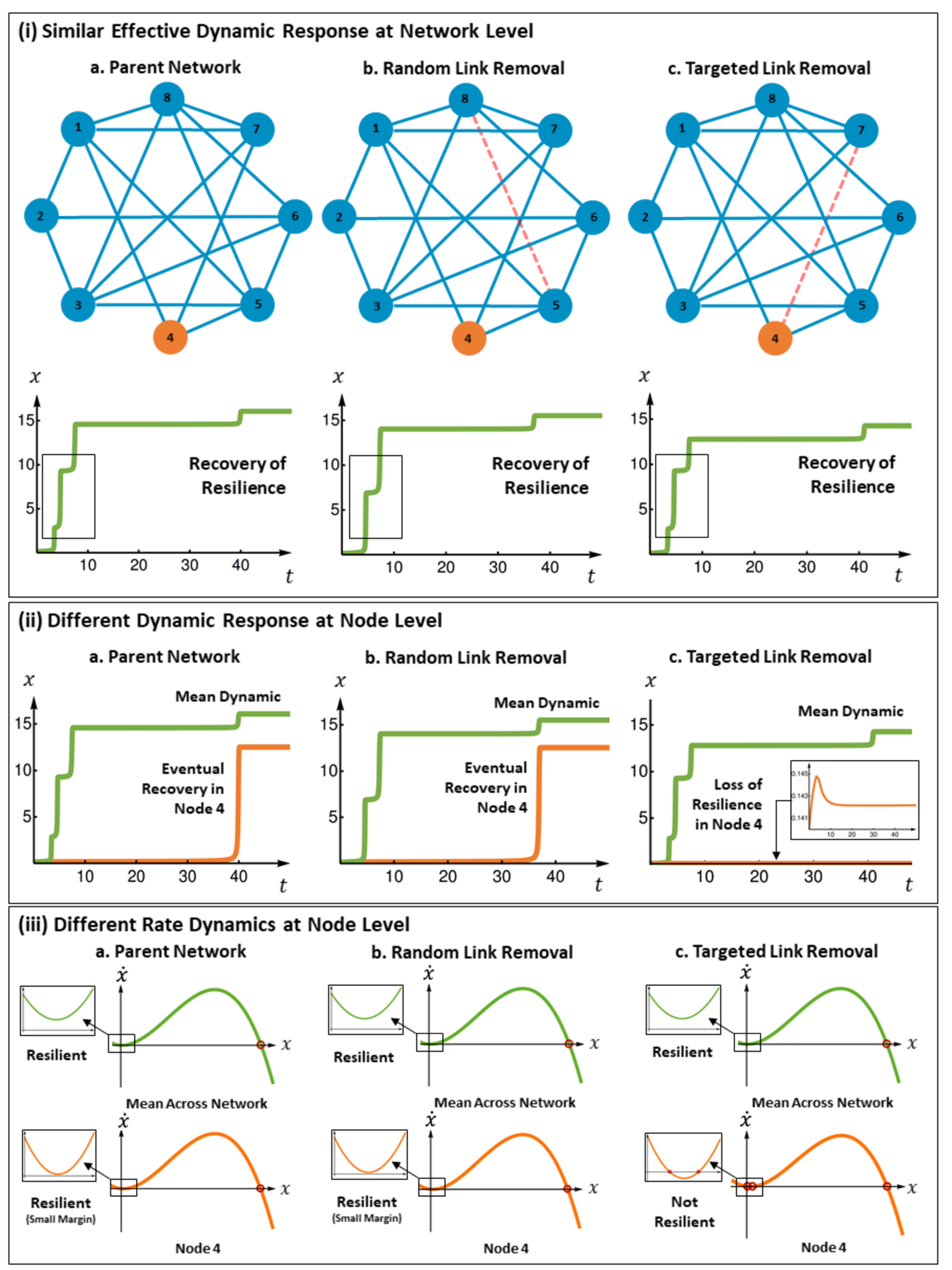}
    \caption{{\bf Similar Networks Dynamics - Different Node Dynamics.} \textbf{(i)} Similar Network Dynamic Response - (a) the parent network can have a link removed either (b) randomly or (c) targeted to cause local resilience loss. At the network level, the effective behaviour (analysis using mean field \cite{Gao16}) is all similar: demonstrating that the whole system's mean behaviour can recover. \textbf{(ii)} Different Node Dynamic Response - However, we show that there is a loss of resilience in node 4 for case (c) by design. Whilst this detail is lost in the network level mean behaviour, it can be predicted using our proposed framework. \textbf{(iii)} Different Node Rate Dynamics - This shows that whilst we retain a similar resilient profile across the network across all cases, we can clearly see that node 4 is marginally above resilience in the parent network, remains resilient after random link removal, but looses resilience after targeted link removal.}
    \label{fig:2}
\end{figure}

\begin{figure*}[t]
    \centering
    \includegraphics[width=0.95\linewidth]{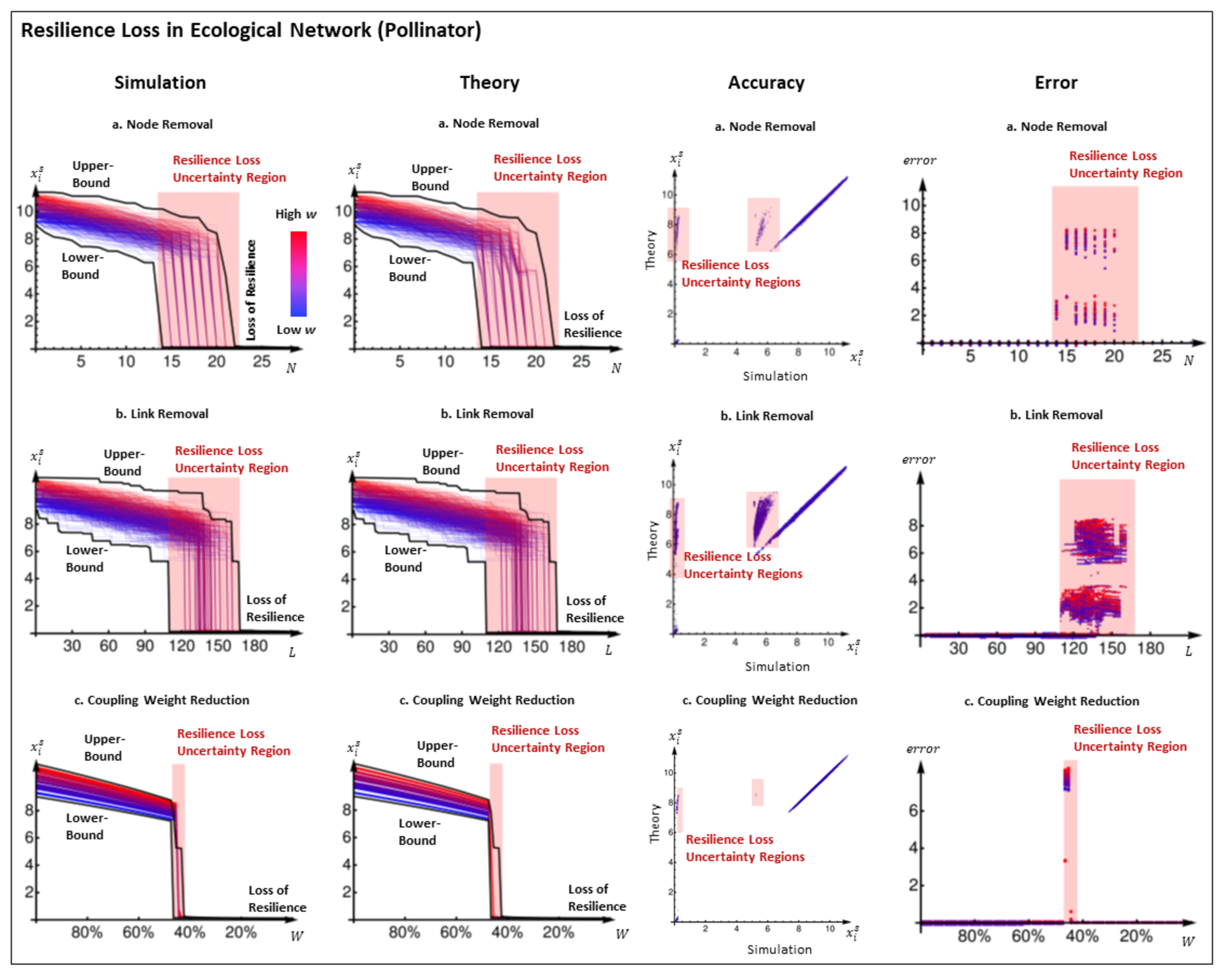}
    \caption{{\bf Resilience Function in Ecological Network (Pollinator).} Resilience Loss from random (a) node removal $N$, (b) link removal $L$, and (c) coupling weight loss $W$. Results show both Monte-Carlo simulated behaviour, theoretical predictions, accuracy of predictions, and error margin outside the uncertain transition region. The estimation steps used is $s=3$.}
    \label{fig:3}
\end{figure*}
\begin{figure*}[t]
    \centering
    \includegraphics[width=0.95\linewidth]{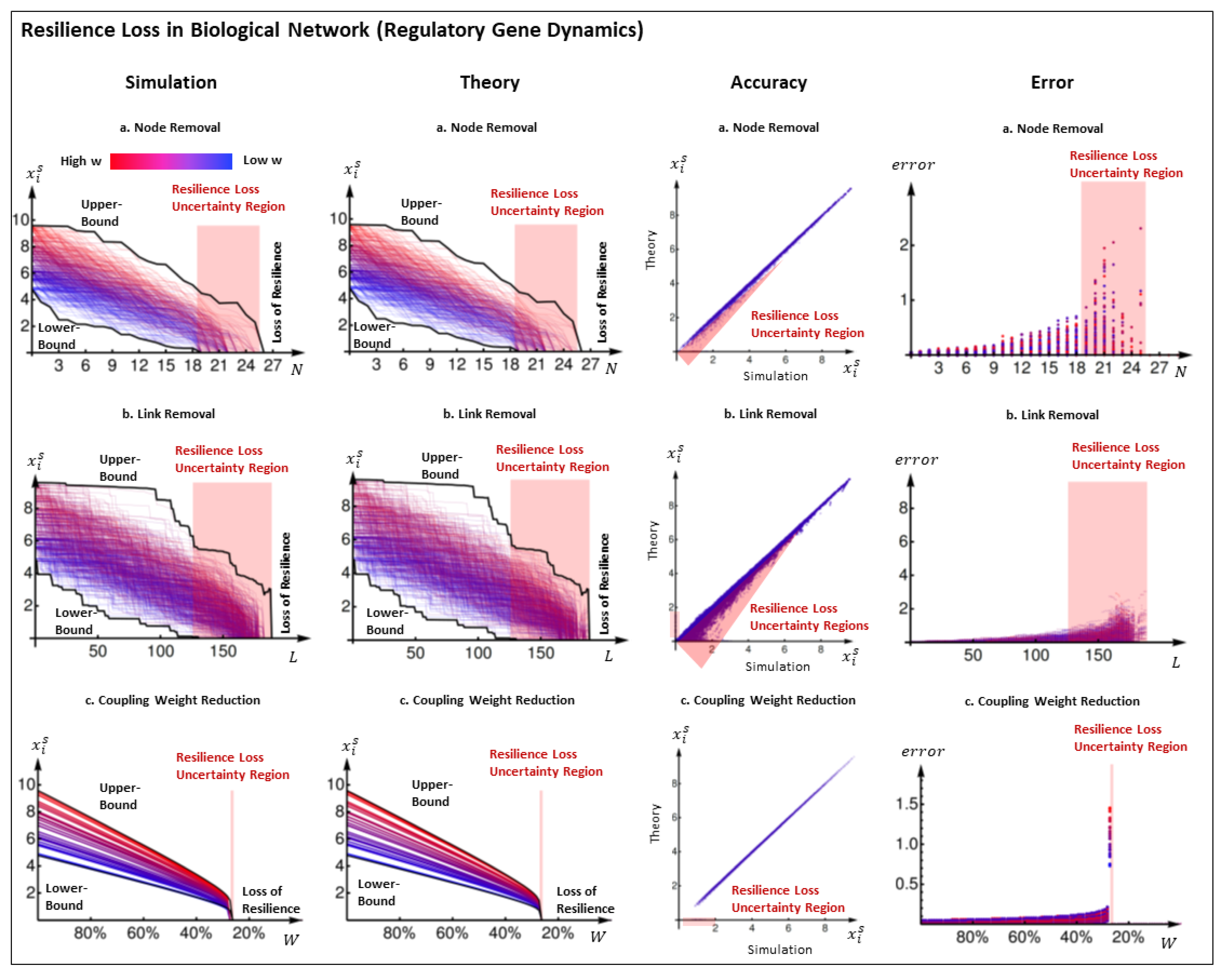}
    \caption{{\bf Resilience Function in Biological Network (Gene Regulation).} Resilience Loss from random (a) node removal $N$, (b) link removal $L$, and (c) coupling weight loss $W$. Results show both Monte-Carlo simulated behaviour, theoretical predictions, accuracy of predictions, and error margin outside the uncertain transition region. The estimation steps used is $s=0$ (1 initial step).}
    \label{fig:4}
\end{figure*}

\subsection{Approach \& Methodology}

To answer this question, this paper presents a sequential estimation approach (\textbf{detailed methodology in SI}). This enables us to understand how network topology affects the resilience function of a node (see Fig.\ref{fig:1}i-c). As an overview to the coupled dynamical system, we show how 3 key characteristic functions map to each other (see Fig.\ref{fig:1}ii). First, we show that the dynamic response of the whole networked system or an individual node, can have a context dependent desirable and undesirable operating state. The dynamic response describes if the system can bounce back from undesirable to a desirable state. Second, we show that the rate dynamics $\dot{x}(x)$ defines the equilibrium states of the system, where there are desirable equilibrium and potentially undesirable ones (formation of which depends on dynamics, network, and perturbations). Finally, we show that by understanding the aforementioned dynamics, we can predict the resilience function of the whole system and each node using the proposed framework.

The proposed framework utilizes an initial homogeneous mean field estimation (Fig.\ref{fig:1}iii) to drive sequential substitution and evaluation of heterogeneous resilience at each node (Fig.\ref{fig:1}iv). 

\textbf{Step 0:} First, we calculate a mean field approximation of the system. By using either a homogeneous average degree $w_{\text{av}} = \frac{1}{N}\sum_{i}^{N}\sum_{j}^{N}a_{ij}$ or a weighted average degree $w_{\text{av}} = \frac{\langle w_{\text{out}}\, w_{\text{in}} \rangle}{\langle w_{\text{out}} \rangle}$ we calculate the equilibrium $\eqapprox{0}$ of the dynamical system
\begin{equation}
\frac{d x}{d t} = f(x) + \waverage g(x,x).
\label{eq:meanfield}
\end{equation} The relative merits of the two way to define $w_{\text{av}}$ are discussed in \textbf{SI}. 

\textbf{Step 1} We use the mean field approximation as an initial guess to bootstrap our approximations. We approximate the dynamics on each node by the dynamical system
\begin{equation}
\frac{d x}{d t} = f(x) + \win_i g(x,\eqapprox0) = 0.
\label{eq:step1}
\end{equation}
This gives us a new value for the equilibrium of the system, $\veqapprox1$. We call this
the first order approximation of the equilibrium.

\textbf{Step 2 to $\mathbf{n}$} We use the previous approximation to approximate the effect that the graph has on a single vertex by looking at the dynamical system
\begin{equation}
\frac{d x}{d t} = f(x) + \win_i \frac{\langle \dout \cdot g(x,\veqapprox{n}) \rangle}{\langle \dout \rangle}.
\label{eq:steps}
\end{equation} In order to find their mean effect of the neighbours, each component of the coupling vector $g(\cdot)$ is weighted by $\dout$ (this is explained in detail in the \textbf{SI}). This gives us an updated estimate of the equilibrium $\veqapprox{n+1}$. We call this the $n+1$-th order approximation of the equilibrium. 

By estimating the equilibrium solutions at each node subject to perturbations, we are able to infer node-level dynamic response and resilience functions. The resulting framework is a robust and accurate way of measuring the networked dynamics and resilience function at each node, with the ability to identify vulnerable nodes. We can generally predict the resilience function with up to 98\% accuracy after $s<2$ steps of estimation. Furthermore, it mathematically links topological measures and non-linear dynamics (relationship shown in Fig.\ref{fig:1}i-b). We demonstrate its capability through commonly studied ecological systems, subject to the standard perturbation models of: (i) node loss, (ii) link loss, and (iii) weight loss. We expect this new and transformative framework will map to existing application domain knowledge and inform the design and operations in a wide range of domains.\\

\section{Results}

\subsection{Node Level Resilience} 

The key benefit of our proposed framework is the ability to identify vulnerable nodes that are at risk of losing resilience. This is done so by examining the impact of perturbations on the effective resilience of the whole network \cite{Gao16}, as well as the individual resilience of nodes. Here, our results in Fig.\ref{fig:2}i show that a parent network (case a) can have a similar effective dynamics after perturbation. In this example we use a random link removal (case b) and a targeted link removal (case c). We see that the network's mean dynamic response to recover a certain desirable equilibrium solution is similar (small differences highlighted in black box). However, when we look at an individual node's dynamic response (node 4 in Fig.\ref{fig:2}ii), we observe 2 effects. First, we see that node 4 recovers its desirable functionality (case a and b) with a longer delay. Second, we see that when targeted link removal (case c) is performed, node 4 never recovers (zoom in shows it collapses to a low equilibrium value). 

These results highlight a shortfall in current approaches that only estimate network level dynamics \cite{Gao16, Tu_PRE_17, Liang17, Jiang_PNAS_17}, whereby all 3 cases have near identical mean field values and as such yield similar mean network dynamics. That means practitioners are unable to identify vulnerabilities at the node level and gain more insight or direct interventions. Explaining the node level results, Fig.\ref{fig:2}iii shows the rate dynamics. Again, a similar network level behaviour exists before and after perturbations. However, at the node 4 level, we can see how targeted link removal can shift it from resilient with a small margin to not resilient. The dynamics used in Fig.\ref{fig:2} are: $f(x) = x_{i}(1-\frac{x_{i}}{5})(x_{i}-1)$, and $g(x_{i},x_{j}) = \frac{x_{i}^{2}}{2(x_{j}+1)}$, and the estimation steps used is $s=1$ (2 steps). Later in the paper, we will present more complex dynamical systems, where the results are less intuitive. For now, to motivate readers, we present 2 case studies motivated by examples given in \cite{Barzel13, Gao16}.

\subsection{Case Study: Ecological Network} 

Here in Fig.\ref{fig:3}, we use a well studied case of pollinator networks \cite{Holland02}. The abundance of species $i$, $x_{i}$ is given by:
\begin{equation}
\frac{dx_{i}}{dt} = x_{i}(1-\frac{x_{i}}{K})(\frac{x_{i}}{C}-1) + \sum_{j}^{N} a_{ji}\frac{x_{i}x_{j}}{D_{i}+E_{i}x_{i}+H_{j}x_{j}} + B_{i},
\label{eq:ecology}
\end{equation} where with reference to Eq.\eqref{eq:fg}, $f(.)$ is a logistic growth equation balancing the carrying capacity $K_{i}$ with the Allee effect (low abundance $x_{i}<C_{i}$ leads to population decline), $g(.)$ in Eq.\eqref{eq:fg} is a coupling function with saturation, and $B_{i}$ is a constant migration rate from other ecosystems. For simplicity, we use homogeneous functionality parameters: $B=0.1, C=1, K=5, D=5, E=0.9, H=0.1$. For topological generality in all our case studies, we used random graphs and in this case it is a Bernoulli graph with $N=30$ nodes and a connectivity factor of $p=0.5$. Other random graphs exhibit similar results. 

Our results in Fig.\ref{fig:3} show that outside the resilience loss regime (red region), we are able to predict well the effect of the different perturbations, with less than 2-4\% error with estimation steps $s=3$ and an initial mean field approximation of $w_{\text{av}}$. From the results, we can see that due to the Allee effect, the collapse in abundance in every species is dramatic after a certain perturbation level. We are able to create upper- and lower-bounds for the dynamics, such that we can estimate the size of the uncertainty region. We can see that within the uncertainty region, the error can be arbitrarily large - highlighting unpredictable behaviour during resilience loss. The impact of this work is that we can clearly predict the onset of resilience loss for different measurable perturbation dynamics. We can see the impact of changing either specific species parameters (e.g. carrying capacity or colony threshold) and overall spatial network level connectivity on the resilience profile of both the whole ecosystem and the specific species. 

\subsection{Case Study: Biological Network} 

Here in Fig.\ref{fig:4}, we use a well studied case of gene regulatory networks governed by the Michaelis-Menten equation \cite{Alon06}, given by:
\begin{equation}
\frac{dx_{i}}{dt} = -x_{i}^{a} + \sum_{j}^{N} a_{ji}\frac{x_{j}^{h}}{2(x_{j}^{h}+1)},
\label{eq:metabolic}
\end{equation} where with reference to Eq.\eqref{eq:fg}, $f(.)$ is a degradation ($a=1$) or dimerization ($a=2$) effect, and $g(.)$ in Eq.\eqref{eq:fg} is genetic activation, where the Hill coefficient $h$ describes the level of cooperation in gene regulation. Using $a=1, h=2$, we find that there is a more gradual loss of resilience than the pollinator network. We show that outside the resilience loss regime (red region), we are able to predict well the effect of perturbations, with an initial error of less than 2\% (rising gradually), with estimation steps $s=0$ (1 initial mean field step, because the dynamics are trivial) and the initial mean field approximation of $\frac{\langle w_{\text{in}}w_{\text{out}} \rangle}{\langle w \rangle}$ used in \cite{Gao16}. We are also able to create upper- and lower-bounds for the dynamics, such that we can estimate the size of the uncertainty region. This case study demonstrates that when the dynamics are relatively trivial (no $x_{i}$ in coupling dynamics $g(.)$), we can predict the gradual resilience loss very well. Later in the next section, we will show how a critical resilience function can be used to identify the most vulnerable nodes and how for non-trivial cases, the resulting resilience functions can be non-intuitive.
\begin{figure}[t]
    \centering
    \includegraphics[width=1.00\linewidth]{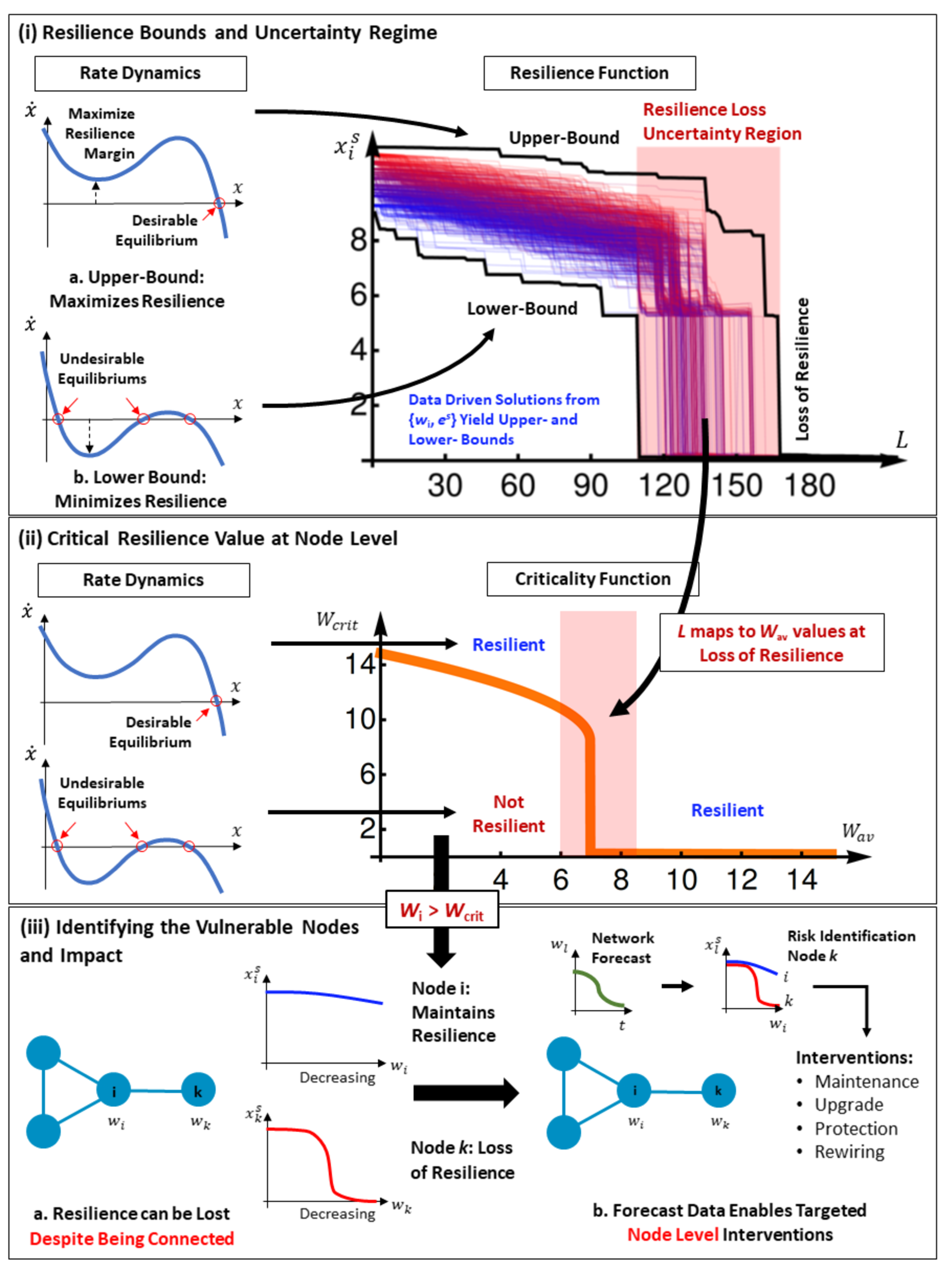}
    \caption{{\bf Critical Resilience Value Identifies Vulnerable Nodes.} \textbf{(i)} Resilience Bounds and Uncertainty Regime - upper-bound maximizes and lower-bound minimizes resilience, whereby real network data and dynamics $(w_{i},e^{n})$ drive the bound form. Resilience loss at upper- and lower-bound maps to average weighted degree values ($w_{\text{av}}$), which map to critical resilience value plot. \textbf{(ii)} Criticality Function defines resilience regimes mapping network properties (average weighted degree values $w_{\text{av}}$) to local node properties (critical resilience value $w_{\text{crit}}$). When $w_{i} > w_{\text{crit}}(w_{\text{av}})$, the node is resilient, and when below it is not resilient. This is a way to identify which nodes are likely to be vulnerable to a loss of resilience. \textbf{(iii)} Impact - This shows that nodes do not have to be removed to lose resilience. By being able to identify and forecast which nodes are at risk of resilience loss as a function of parameters (e.g. declining interactions $w$ over time), we can target interventions for different contexts.}
    \label{fig:5}
\end{figure}

\subsection{Resilience Bounds and Critical Resilience Value} 

The crux of our work is to tractably analyze node level resilience and use this to identify which nodes are at risk of loosing resilience. In Fig.\ref{fig:5}, we use the pollinator dynamics (see Eq.\eqref{eq:ecology}) to demonstrate how to estimate the upper- and lower-bounds of the resilience function and identify vulnerable nodes. For this particular dynamic, Fig.\ref{fig:5}i shows how the upper- and lower-bounds can be found by maximizing and minimizing the rate dynamics. The bound solutions are subject to the real data available on both the topology ($w_{i}$) and the equilibrium estimation $e^{s}$. When the bounds collapse, we can use their $w_{\text{av}}$ values to map the uncertainty regime to other relationship plots. One such plot is the criticality function. Here, we define the critical resilience $w_{\text{crit}}$ as the value by which each node must satisfy $w_{i} > w_{\text{crit}}$ in order to stay resilient. This condition enables us to identify which nodes maybe close to losing resilience - see Fig.\ref{fig:2}iii, despite being reasonably well connected, and use future knowledge of connectivity changes to drive forecasting of resilience loss at the node level.
\begin{figure}[t]
    \centering
    \includegraphics[width=1.00\linewidth]{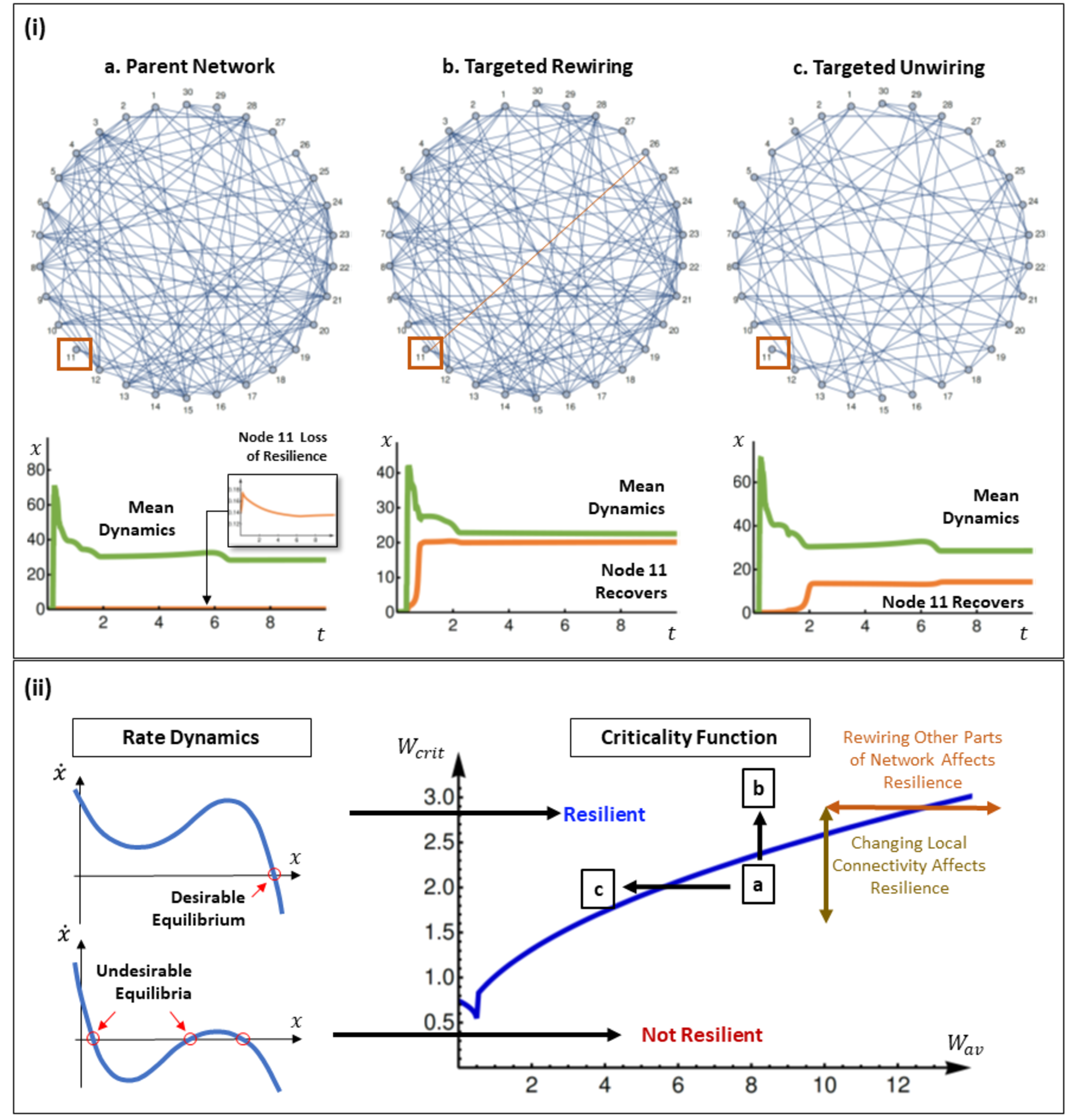}
    \caption{{\bf Network Rewiring \& Unwiring to Change Resilience.} \textbf{(i)} (a) parent network's node 11 has failed, (b) targeted rewiring by adding an edge allows node 11 to recover, (c) targeted unwiring of other parts of the network allows node 11 to recover. \textbf{(ii)} criticality function with inflection point, showing that rewiring/unwiring one part of the network has different effects on the resilience for other parts of the network. In our example, the parent network's node 11 moved from a not resilient regime (a) to a resilient regime by increasing its own connectivity (b) or unwiring other network parts (c).}
    \label{fig:6}
\end{figure}

\subsection{Network Rewiring \& Unwiring} 

Thus far, the dynamics employed have been motivated by real ecosystem and biological system cases. Due to the relative simplicity of the coupling dynamics $g(x_i,x_j)$, the identification of the most vulnerable nodes at risk of resilience loss are in some cases intuitive. For example, in the case of pollinator dynamics, stronger coupling will improve resilience. We now show that changes to the coupling dynamics can dramatically change the results. We consider the following dynamics: $f(x_i) = 1/10 + x_{i}(1-x_{i})(x_{i}/5-1)$ and $g(x_{i},x_{j}) = 15 (x_{i}^{2}x_{j})/(1 + x_{j}^2)$. In this case, increasing the connectivity of one part of the network can have opposite effects on the resilience of other nodes, depending on their local connectivity and global network topology. 

In Fig.\ref{fig:6}i, we consider a parent network (case a), whereby node 11 has already lost resilience - see dynamics $x(t)$. In Fig.\ref{fig:6}ii, the position of node 11 on the criticality function is labelled and is in a not resilient regime. When we perform targeted rewiring (case b), node 11 is connected to node 26 and its dynamic response recovers. In In Fig.\ref{fig:6}ii, rewiring improves its local weighted degree $w_{11} > w_{\text{crit}}$ and shifts the position upwards into a resilient regime. When we perform targeted unwiring (case c), several links are removed elsewhere in the network and this reduces the average weighted degree $w_{\text{av}}$ such that it shifts its position leftwards into a resilient regime. In summary, it is entirely plausible to have a system whereby its dynamics makes intuitive analysis impossible. In such cases, making a small change in one part of the network can dramatically improve resilience for some nodes, whilst reducing resilience for others, depending on where they are with respect to critical inflection point in Fig.\ref{fig:6}ii. The fact that local changes can affect resilience in a completely different part of the network deserves attention and further research. \\

\section{Discussion \& Limitations}

A gap in understanding exists between individual dynamics and the coupled dynamics in a large-scale networked complex system. Here, we present a framework for tractably analyzing the resilience of individual nodes as a function of the individual dynamics and the network property. We show that it can estimate: (1) the equilibrium behaviour outside the critical region, (2) estimate the critical region as a function of the perturbation, (3) identify the nodes that are most vulnerable to loss of resilience, and (4) predicting the effect of changing the network on the resilience of nodes. Whilst our baseline result is intuitive (e.g. the most vulnerable nodes are poorly connected ones close to the critical resilience value $w_{\text{crit}}$), quantifying this value as a function of the dynamics enables us to prioritize actions more effectively and predict resilience loss more accurately. Conversely, we may also discover hidden cascade effects, whereby disconnecting a weakly connected node can lead to failure in other nodes. For example, recent claims on eradicating the malaria mosquito because it is not a significant diet for predators (e.g. weak basal species \cite{Collins18}) maybe risky, because we do not know the underlying dynamics nor the resilience margins in all species. 

It is also useful to discuss when our estimation algorithm doesn't work. As with \cite{Gao16}, the estimation produces increasing errors with (1) increasing degree correlation (assortiveness) and (2) clustering coefficient. This is likely due to the graph containing a mixture of topological distributions, which makes mean field approximations less accurate. Putting this caveat aside, in general, the node-level precision methods developed here will enable practitioners in ecology, infrastructure, and other application areas to prioritize protection and intervention resources, such as maintenance, preservation, rewiring, and upgrades. Future work will extend this research to consider both local \cite{Klickstein17} and global \cite{Gao17} optimal control of complex network dynamics. \\

\bibliographystyle{IEEEtran}
\bibliography{IEEEabrv,main}

\vspace{5mm}

\textbf{Acknowledgments}
W.G. and G.M. conceived the idea, designed the analysis and wrote the paper. G.M. sourced the data and conducted the analysis. 

The authors have no competing financial interests to report. \\
Correspondence and requests for materials should be addressed to: weisi.guo@warwick.ac.uk \\
Email address of G. Moutsinas: g.moutsinas.1@warwick.ac.uk \\

\clearpage

\end{document}